\documentclass[12pt,english]{article}
\usepackage[latin9]{inputenc}
\usepackage{geometry}
\usepackage{babel}
\usepackage{amsmath}
\usepackage{graphicx}
\usepackage{setspace}
\usepackage[authoryear]{natbib}
\usepackage[unicode=true,pdfusetitle,
 bookmarks=true,bookmarksnumbered=false,bookmarksopen=false,
 breaklinks=false,pdfborder={0 0 1},backref=false,colorlinks=false]
 {hyperref}

\makeatletter
\newcommand{\lyxaddress}[1]{
\par {\raggedright #1
\vspace{1.4em}
\noindent\par}
}

\@ifundefined{date}{}{\date{}}
\makeatother

\begin{document}

\title{Statistics of spatial averages and optimal averaging in the presence
of missing data}

\author{Ashwin K Seshadri}
\maketitle

\lyxaddress{Divecha Centre for Climate Change, Indian Institute of Science, Bangalore,
India, 560012. Email: ashwin@fastmail.fm\pagebreak{}}
\begin{abstract}
We consider statistics of spatial averages estimated by weighting
observations over an arbitrary spatial domain using identical and
independent measuring devices, and derive an account of bias and variance
in the presence of missing observations. We test the model relative
to simulations, and the approximations for bias and variance with
missing data are shown to compare well even when the probability of
missing data is large. Previous authors have examined optimal averaging
strategies for minimizing bias, variance and mean squared error of
the spatial average, and we extend the analysis to the case of missing
observations. Minimizing variance mainly requires higher weights where
local variance and covariance is small, whereas minimizing bias requires
higher weights where the field is closer to the true spatial average.
Missing data increases variance and contributes to bias, and reducing
both effects involves emphasizing locations with mean value nearer
to the spatial average. The framework is applied to study spatially
averaged rainfall over India. We use our model to estimate standard
error in all-India rainfall as the combined effect of measurement
uncertainty and bias, when weights are chosen so as to yield minimum
mean squared error. 
\end{abstract}

\section{Introduction}

Spatial averages occur frequently in climate science, for example
in global or regional temperature and precipitation (\citet{Vinnikov1990}),
and these are estimated by averaging across what are often point observations.
The weights used in spatial averaging do not have to be uniform, even
if observations themselves are uniformly spaced, but can differ according
to the goal of the inquiry (\citet{Vinnikov1990,Gandin1993,Benedetti1995}).
In applications involving statistics, not only point estimates of
the spatial average but also knowledge of additional quantities such
as the bias or variance becomes important (\citet{Cressie1991,Morrissey1999,Casella2002}).
An old problem in spatial statistics is optimal averaging (OA), where
investigators have examined the choice of weights minimizing bias,
random error, or mean squared error (MSE) in estimates of the spatial
average of a geophysical quantity (\citet{Vinnikov1990,Gandin1993,Shen1994,Overton1993,Benedetti1995,Shen1998,Morrissey1999,Vinnikov1999}). 

A common feature of this previous literature (for e.g. \citet{Vinnikov1990,Gandin1993,Shen1994,Morrissey1999})
is the assumption that the relevant observations would all be reported.
However the general case admits situations where some observations
are missing, so that the average can only be evaluated over that part
of the domain for which measurements are available. Nevertheless such
situations may require prior estimates of variance or MSE, and averaging
strategies must consider the goal of averaging in light of the ensemble
of possible realizations over missing data. With the possibility of
missing data the spatial average can be estimated as the ratio between
linearly weighted observations over the domain and the fraction of
the domain represented by available observations. The present paper
derives estimators for bias and variance of such a spatial average
in the presence of missing data. 

In contrast to the case where all measurements are reported, the possibility
of missing data induces variability in the denominator as well as
covariance between the numerator and denominator, both of which must
additionally be estimated. Statistics of a ratio between two quantities
do not yield exact expressions, and one must resort to approximate
methods (\citet{Hartley1954,Oehlert1992}). We estimate statistics
(squared bias, variance, and MSE) based on truncation of a Taylor
expansion of the ratio (\citet{Oehlert1992,Vaart1998}) and examine
the accuracy of resulting approximations. 

Missing data introduces new features, and increases the bias and variance
in the spatial average. The model developed here helps understand
these effects. This is applied to examine statistics of rainfall averaged
over India (\citet{Mooley1984,Gadgil2003}). The model of variance
is used to describe temporal variability of all-India rainfall. In
addition we estimate optimal weighs that would minimize MSE in estimates
of all-India rainfall. In statistical applications, for example involving
variability and change (\citet{Mooley1984}), involving estimates
of spatially averaged quantities it is necessary to know the standard
error associated with the estimates (\citet{Nicholls2014}). Here
we provide a rough quantification of standard error associated with
estimates of all-India rainfall. The standard error measures uncertainty
associated with reports of the all-India average, and arises from
the the total contribution to MSE from measurement uncertainty and
squared bias. 

Our essential problem is as follows. Consider point observations $r_{i}$
of spatially-varying field $\upsilon$ over a domain at locations
indexed by $i$ $\left(1\leq i\leq n\right)$ and corresponding to
a fixed interval of time. Individual observations are assigned fixed
weights $\beta_{i}$, and relatively unbiased estimation of the spatial
average requires $\sum_{i}\beta_{i}=1$. Some observations might not
be reported when sought, and Boolean random variable $s_{i}$ represents
this status: if available then $s_{i}=1$ otherwise $s_{i}=0$. Availability
of observation at a point does not impinge on availability at another,
so the $s_{i}$'s are statistically independent. If measuring and
reporting instruments are nearly identical in this aspect, then each
$s_{i}$ has known probability $\alpha$ of equaling one when called
for. 

We examine statistics of spatial average 
\begin{equation}
r\equiv\frac{R}{S}=\frac{\sum_{i}\beta_{i}s_{i}r_{i}}{\sum_{i}\beta_{i}s_{i}}\label{eq:n1}
\end{equation}
which merely estimates the true spatial average $\upsilon$. Generally,
due to inherent variability, both $\upsilon$ and $r$ are functions
of time. Point observation $r_{i}$ is assumed to be related to the
corresponding true value $\upsilon_{i}$ through additive noise. 

Section 2 derives approximate estimators for squared bias, variance
and MSE of the above estimator $r$, assuming that $\beta_{i}$'s
are known. Section 3 examines how to decide weights $\beta_{i}$ for
minimizing the chosen statistic, given knowledge of the statistics
of observations as well as probability $\alpha$ of individual observations
being available. Section 4 illustrates for the case of all-India rainfall.
We illustrate major effects, including that of missing data, and consider
an application: estimation of the standard error in the spatial average
of rainfall. 

\section{Statistics of a spatial average}

\subsection{Approximate statistics through \char`\"{}Delta-method\char`\"{}}

The spatial average is is denoted as a function $r=f\left(R,S\right)$
of two variables, and its Taylor series about $\left(\boldsymbol{\mathbf{E}}R,\boldsymbol{\mathbf{E}}S\right)$,
with $\boldsymbol{\mathbf{E}}$ denoting expectation, truncated to
2\textsuperscript{nd} degree is
\begin{multline}
f\left(R,S\right)=f\left(\boldsymbol{\mathbf{E}}R,\boldsymbol{\mathbf{E}}S\right)+\left(R-\boldsymbol{\mathbf{E}}R\right)\frac{\partial f}{\partial R}+\left(S-\boldsymbol{\mathbf{E}}S\right)\frac{\partial f}{\partial S}+\frac{1}{2}\left(R-\boldsymbol{\mathbf{E}}R\right)^{2}\frac{\partial^{2}f}{\partial R^{2}}+\frac{1}{2}\left(S-\boldsymbol{\mathbf{E}}S\right)^{2}\frac{\partial^{2}f}{\partial S^{2}}+\\
\left(R-\boldsymbol{\mathbf{E}}R\right)\left(S-\boldsymbol{\mathbf{E}}S\right)\frac{\partial^{2}f}{\partial R\partial S}\label{eq:p2}
\end{multline}
with partial derivatives evaluated at $\left(\boldsymbol{\mathbf{E}}R,\boldsymbol{\mathbf{E}}S\right)$
being $\frac{\partial f}{\partial R}=\frac{1}{\boldsymbol{\mathbf{E}}S}$,
$\frac{\partial f}{\partial S}=-\frac{\boldsymbol{\mathbf{E}}R}{\left(\boldsymbol{\mathbf{E}}S\right)^{2}}$,
$\frac{\partial^{2}f}{\partial R^{2}}=0$, $\frac{\partial^{2}f}{\partial S^{2}}=\frac{2\boldsymbol{\mathbf{E}}R}{\left(\boldsymbol{\mathbf{E}}S\right)^{3}}$,
$\frac{\partial^{2}f}{\partial R\partial S}=-\frac{1}{\left(\boldsymbol{\mathbf{E}}S\right)^{2}}$
, and taking the expectation
\begin{equation}
\boldsymbol{\mathbf{E}}f\left(R,S\right)=\frac{\boldsymbol{\mathbf{E}}R}{\boldsymbol{\mathbf{E}}S}+\frac{\boldsymbol{\mathbf{E}}R}{\left(\boldsymbol{\mathbf{E}}S\right)^{3}}\boldsymbol{\mathbf{E}}\left(S-\boldsymbol{\mathbf{E}}S\right)^{2}-\frac{1}{\left(\boldsymbol{\mathbf{E}}S\right)^{2}}\boldsymbol{\mathbf{E}}\left(R-\boldsymbol{\mathbf{E}}R\right)\left(S-\boldsymbol{\mathbf{E}}S\right)\label{eq:p3}
\end{equation}
the assumption being, of course, that $f\left(R,S\right)$ is smooth
in the neighborhood of $f\left(\boldsymbol{\mathbf{E}}R,\boldsymbol{\mathbf{E}}S\right)$
so that locally it can be approximated by the first few terms. This
approach, sometimes called the \char`\"{}delta method\char`\"{}, approximates
the expectation of a function by that of its Taylor series and converges
if the function is sufficiently smooth and has finite moments (\citet{Oehlert1992,Vaart1998}).
The true spatial average is $\upsilon$, and the MSE in estimating
it is 
\begin{equation}
\mathbf{MSE}=\boldsymbol{\mathbf{E}}\left(f\left(R,S\right)-\upsilon\right)^{2}=\boldsymbol{\mathbf{E}}\left(f\left(R,S\right)-\boldsymbol{\mathbf{E}}f\left(R,S\right)\right)^{2}+\boldsymbol{\mathbf{E}}\left(\boldsymbol{\mathbf{E}}f\left(R,S\right)-\upsilon\right)^{2}\label{eq:p4}
\end{equation}
being described the sum of variance $V_{r}=\boldsymbol{\mathbf{E}}\left(f\left(R,S\right)-\boldsymbol{\mathbf{E}}f\left(R,S\right)\right)^{2}$
and squared bias $B_{r}=\boldsymbol{\mathbf{E}}\left(\boldsymbol{\mathbf{E}}f\left(R,S\right)-\upsilon\right)^{2}$,
whose derivation is standard and therefore omitted. Henceforth we
shall refer to $B_{r}$ as simply bias. 

In case there is no possibility of missing observations, then $S$
is fixed and $\boldsymbol{\mathbf{E}}\left(S-\boldsymbol{\mathbf{E}}S\right)^{2}$,
$\boldsymbol{\mathbf{E}}\left(R-\boldsymbol{\mathbf{E}}R\right)\left(S-\boldsymbol{\mathbf{E}}S\right)$,
etc., vanish. In that case $S=\sum_{i}\beta_{i}=1$ , and the estimator
of the spatial average is simply $R=\sum_{i}\beta_{i}r_{i}$. Its
variance simplifies to $V_{r}=\boldsymbol{\mathbf{E}}R^{2}-\left(\boldsymbol{\mathbf{E}}R\right)^{2}$
and bias $B_{r}=\boldsymbol{\mathbf{E}}\left(\boldsymbol{\mathbf{E}}R-\upsilon\right)^{2}$.
Now, using linearity of $\mathbf{E}$, $\mathbf{E}R^{2}=\sum_{i}\beta_{i}^{2}\mathbf{E}r_{i}^{2}+2\sum_{i<j}\beta_{i}\beta_{j}\mathbf{E}r_{i}r_{j}$,
and $\left(\mathbf{E}R\right)^{2}=\sum_{i}\beta_{i}\left(\mathbf{E}r_{i}\right)^{2}+2\sum_{i<j}\beta_{i}\beta_{j}\mathbf{E}r_{i}\mathbf{E}r_{j}$
so that 
\begin{equation}
V_{r}=\sum_{i}\beta_{i}^{2}\sigma_{r_{i}}^{2}+2\sum_{i<j}\beta_{i}\beta_{j}\mathbf{Cov}\left(r_{i},r_{j}\right)\label{eq:n5}
\end{equation}
where $\mathbf{Cov}\left(r_{i},r_{j}\right)$ is the covariance. This
formula can be depicted as positive-definite quadratic form $V_{r}=\beta^{T}S_{r}\beta$
where $S_{r}$ is the covariance matrix of observations and $\beta=\left\{ \begin{array}{ccc}
\beta_{1} & \ldots & \beta_{n}\end{array}\right\} ^{T}$ is the vector of weights. Similarly the squared bias reduces to $\boldsymbol{\mathbf{E}}\left(\sum_{i}\beta_{i}\boldsymbol{\mathbf{E}}r_{i}-\upsilon\right)^{2}$
which, given that the weights sum to unity, is estimated by average
across time-series $\frac{1}{N}\sum_{t}\left(\sum_{i}\beta_{i}\left(\boldsymbol{\mathbf{E}}r_{i}\left(t\right)-\upsilon\left(t\right)\right)\right)^{2}$
. Defining vector $d_{1}\left(t\right)=\left\{ \begin{array}{cccc}
\boldsymbol{\mathbf{E}}r_{1}\left(t\right)-\upsilon\left(t\right) & \boldsymbol{\mathbf{E}}r_{2}\left(t\right)-\upsilon\left(t\right) & \ldots & \boldsymbol{\mathbf{E}}r_{n}\left(t\right)-\upsilon\left(t\right)\end{array}\right\} ^{T}$, we obtain $B_{r}=\frac{1}{N}\sum_{t}\beta^{T}d_{1}\left(t\right)d_{1}\left(t\right)^{T}\beta$
or equivalently 
\begin{equation}
B_{r}=\beta^{T}D_{1}\beta\label{eq:n6}
\end{equation}
where $D_{1}=\frac{1}{N}\sum_{t}d_{1}\left(t\right)d_{1}\left(t\right)^{T}$
is an $n\times n$ matrix and $N$ is the total number of periods
indexed by $t$. These results correspond to those derived by previous
authors for a situation with no missing data (\citet{Vinnikov1990,Shen1998,Vinnikov1999,Shen2007}).

We now return to the general situation where individual observations
are missing with probability $1-\alpha$. For estimating variance,
we approximate $\boldsymbol{\mathbf{E}}f\left(R,S\right)\cong\frac{\boldsymbol{\mathbf{E}}R}{\boldsymbol{\mathbf{E}}S}$
because the other terms are relatively small in case $1-\alpha\ll1$
(Appendix 1), so that variance becomes $\sigma_{r}^{2}=\boldsymbol{\mathbf{E}}\left(f\left(R,S\right)-\frac{\boldsymbol{\mathbf{E}}R}{\boldsymbol{\mathbf{E}}S}\right)^{2}$,
and using the 1\textsuperscript{st} order approximation of $f\left(R,S\right)$
the variance is
\begin{equation}
V_{r}=\boldsymbol{\mathbf{E}}\left\{ \left(R-\boldsymbol{\mathbf{E}}R\right)\frac{\partial f}{\partial R}+\left(S-\boldsymbol{\mathbf{E}}S\right)\frac{\partial f}{\partial S}\right\} ^{2}\label{eq:n7}
\end{equation}
simplifying to
\begin{equation}
\sigma_{r}^{2}=\frac{1}{\left(\boldsymbol{\mathbf{E}}S\right)^{2}}\boldsymbol{\mathbf{E}}\left(R-\boldsymbol{\mathbf{E}}R\right)^{2}+\frac{\left(\boldsymbol{\mathbf{E}}R\right)^{2}}{\left(\boldsymbol{\mathbf{E}}S\right)^{4}}\boldsymbol{\mathbf{E}}\left(S-\boldsymbol{\mathbf{E}}S\right)^{2}-2\frac{\boldsymbol{\mathbf{E}}R}{\left(\boldsymbol{\mathbf{E}}S\right)^{3}}\boldsymbol{\mathbf{E}}\left(R-\boldsymbol{\mathbf{E}}R\right)\left(S-\boldsymbol{\mathbf{E}}S\right)\label{eq:n8}
\end{equation}
or equivalently
\begin{equation}
\sigma_{r}^{2}=\frac{\sigma_{R}^{2}}{\mu_{S}^{2}}+\frac{\mu_{R}^{2}}{\mu_{S}^{4}}\sigma_{S}^{2}-2\frac{\mu_{R}}{\mu_{S}^{3}}\sigma_{RS}^{2}\label{eq:n9}
\end{equation}
where $\mu$ and $\sigma^{2}$ denote denote means and standard deviations
(or covariance) of the subscripted variables. Likewise 
\begin{equation}
\boldsymbol{\mathbf{E}}f\left(R,S\right)-\upsilon=\left\{ \frac{\mu_{R}}{\mu_{S}}-\bar{\upsilon}\right\} +\left\{ \frac{\mu_{R}}{\mu_{S}^{3}}\sigma_{S}^{2}-\frac{\sigma_{RS}^{2}}{\mu_{S}^{2}}\right\} \label{eq:n10}
\end{equation}

with bias being the expectation of the square of this quantity. 

\subsection{Evaluation of the statistics}

For the general situation, where individual observations are missing
with probability $1-\alpha$, the variance of $R$ 
\begin{equation}
\sigma_{R}^{2}\equiv\boldsymbol{\mathbf{E}}\left(R-\mathbf{E}R\right)^{2}=\mathbf{E}R^{2}-\left(\mathbf{E}R\right)^{2}\label{eq:n11}
\end{equation}
is derived in Appendix 2, with result
\begin{equation}
\sigma_{R}^{2}=\alpha\sum_{i}\beta_{i}^{2}\left\{ \mathbf{E}r_{i}^{2}-\alpha\left(\mathbf{E}r_{i}\right)^{2}\right\} +2\alpha^{2}\sum_{i<j}\beta_{i}\beta_{j}\mathbf{Cov}\left(r_{i},r_{j}\right)\label{eq:n12}
\end{equation}
and similarly the variance of $S$ is
\begin{equation}
\sigma_{S}^{2}=\alpha\left(1-\alpha\right)\sum_{i}\beta_{i}^{2}\label{eq:n13}
\end{equation}
arising from uncertainty about whether observations are recorded.
It is largest for $\alpha=0.5$, increasing with uncertainty about
the availability of observations. Covariance between $R$ and $S$
is 
\begin{equation}
\sigma_{RS}^{2}=\alpha\left(1-\alpha\right)\sum_{i}\beta_{i}^{2}\mathbf{E}r_{i}\label{eq:n14}
\end{equation}

Random variable $r_{i}$ describing observation at the $i$\textsuperscript{th}
location is modeled in relation to true value $\upsilon_{i}$ as 
\begin{equation}
r_{i}=\upsilon_{i}+\varepsilon_{i}\label{eq:n15}
\end{equation}
where $\varepsilon_{i}$ is additive noise in the measuring and reporting
instrument. We assume noise has zero mean, i.e. $\mathbf{E}\varepsilon_{i}=0$
, and that $\varepsilon_{i}$ is independent of $\upsilon_{i}$. Hence
\begin{equation}
\mathbf{E}r_{i}^{2}=\mathbf{E}\upsilon_{i}^{2}+E\mathit{\varepsilon_{i}^{2}+2\mathbf{E}\upsilon_{i}\varepsilon_{i}}=\mathbf{E}\upsilon_{i}^{2}+\sigma_{\varepsilon}^{2}\label{eq:n16}
\end{equation}
where $\sigma_{\varepsilon}^{2}$ is the variance of $\varepsilon_{i}$,
independent of $i$ because measuring instruments are assumed identical
in this aspect. The last step used $\mathit{\mathbf{E}\upsilon_{i}\varepsilon_{i}=\mathbf{E}\upsilon_{i}\mathbf{E}\varepsilon_{i}}$
(from independence) and $\mathbf{E}\varepsilon_{i}=0$. Therefore
the expectation of $r_{i}$ is
\begin{equation}
\mathbf{E}r_{i}=\mathbf{E}\upsilon_{i}\label{eq:n17}
\end{equation}
and
\begin{equation}
\mathbf{Cov}\left(r_{i},r_{j}\right)=\mathbf{Cov}\left(\upsilon_{i},\upsilon_{j}\right)\label{eq:n18}
\end{equation}
using Eq.  (\ref{eq:n17}), independence between $\upsilon_{i}$ and
$\varepsilon_{j}$, and assuming noise terms to be mutually independent
($\mathbf{E}\varepsilon_{i}\varepsilon_{j}=\mathbf{E}\varepsilon_{i}\mathbf{E}\varepsilon_{j}=0$).
Substituting Eqs. (\ref{eq:n16}), (\ref{eq:n17}), and (\ref{eq:n18})
into Eq.  (\ref{eq:n12}) yields the variance of $R$
\begin{equation}
\sigma_{R}^{2}=\alpha\sum_{i}\beta_{i}^{2}\left\{ \mathbf{E}\upsilon_{i}^{2}-\alpha\left(\mathbf{E}\upsilon_{i}\right)^{2}\right\} +2\alpha^{2}\sum_{i<j}\beta_{i}\beta_{j}\mathbf{Cov}\left(\upsilon_{i},\upsilon_{j}\right)+\alpha\sigma_{\varepsilon}^{2}\sum_{i}\beta_{i}^{2}\label{eq:n19}
\end{equation}
The first term is related to variances of true values $\upsilon_{i}$.
We define random variable
\begin{equation}
\varsigma_{\upsilon_{i}}^{2}=\mathbf{E}\upsilon_{i}^{2}-\alpha\left(\mathbf{E}\upsilon_{i}\right)^{2}\label{eq:n20}
\end{equation}
which equals the variance of $\upsilon_{i}$ only if $\alpha=1$,
with measurements available with certainty. In general $\varsigma_{\upsilon_{i}}^{2}>\sigma{}_{\upsilon_{i}}^{2}$
because of uncertainty about whether measurements would be available
when called for. The second term in Eq.  (\ref{eq:n19}) arises from
spatial covariance, and the last from measurement uncertainty described
by noise variance $\sigma_{\varepsilon}^{2}$. Similarly, covariance
between $R$ and $S$ becomes, in terms of the field $\upsilon_{i}$
\begin{equation}
\sigma_{RS}^{2}=\alpha\left(1-\alpha\right)\sum_{i}\beta_{i}^{2}\mathbf{E}\upsilon_{i}\label{eq:n21}
\end{equation}
depending on its expected values at the sampled locations. The expectation
of $R$ is $\mu_{R}=\alpha\sum_{i}\beta_{i}\mathbf{E}\upsilon_{i}$
using linearity of $\mathbf{E}$, independence between $r_{i}$ and
$s_{i}$, and Eq.  (\ref{eq:n17}). Similarly that of $S$ is $\mu_{S}=\alpha$.
Substituting Eqs. (\ref{eq:n13}) and (\ref{eq:n14}) into (\ref{eq:n10}),
we have 
\begin{equation}
\boldsymbol{\mathbf{E}}f\left(R,S\right)-\upsilon=\sum_{i}\beta_{i}\mathbf{E}\upsilon_{i}-\upsilon+\left(\frac{1-\alpha}{\alpha}\right)\left\{ \mathtt{\mu_{\upsilon}}\sum_{i}\beta_{i}^{2}-\sum_{i}\beta_{i}^{2}\mathbf{E}\upsilon_{i}\right\} \label{eq:p22}
\end{equation}
with $\mu_{\upsilon}\equiv\sum_{i}\beta_{i}\mathbf{E}\upsilon_{i}$
. Taking the square of the expectation of this quantity yields the
squared bias $B_{r}$. The first contribution $\sum_{i}\beta_{i}\mathbf{E}\upsilon_{i}-\upsilon$
arises from finite sampling of a continuously varying field, and the
second from the possibility of missing observations in case $\alpha<1$. 

As for variance, upon substituting Eqs. (\ref{eq:n13}), (\ref{eq:n14})
and (\ref{eq:n19}) into Eq.  (\ref{eq:n9}) yields 
\begin{multline}
V_{r}=\frac{1}{\alpha}\sum_{i}\beta_{i}^{2}\varsigma_{\upsilon_{i}}^{2}+2\sum_{i<j}\beta_{i}\beta_{j}\mathbf{Cov}\left(\upsilon_{i},\upsilon_{j}\right)+\frac{1}{\alpha}\sigma_{\varepsilon}^{2}\sum_{i}\beta_{i}^{2}-2\left(\frac{1-\alpha}{\alpha}\right)\mu_{\upsilon}\sum_{i}\beta_{i}^{2}\mathbf{E}\upsilon_{i}+\left(\frac{1-\alpha}{\alpha}\right)\mu_{\upsilon}^{2}\sum_{i}\beta_{i}^{2}\label{eq:p23}
\end{multline}
The first three terms in Eq.  (\ref{eq:p23}) are due to variance
in $R$. The fourth term owes to covariance between $R$ and $S$,
and the last to variance in $S$. Generally these statistics involve
higher powers of $\beta_{i}$, and only if $\alpha=1$ do the bias
and variance reduce to quadratic forms. We also recall the use of
only a $1$\textsuperscript{st} order approximation to the spatial
average $f\left(R,S\right)$ in Eq. (\ref{eq:n7}) for computing the
variance. Considering $2$\textsuperscript{nd} and higher order terms,
would have led to higher moments of $R$ and $S$ appearing in our
formula. 

\section{Optimal averaging}

Having derived approximate formulas for bias and variance, let us
consider how to determine the values of $\beta_{i}$'s that would
minimize the chosen statistic. This has been called the optimal averaging
(OA) problem (\citet{Vinnikov1990,Shen1994,Shen1998,Vinnikov1999}).
The novelty in the present work is extension to the case where observations
can be missing with probability $\alpha<1$. 

\subsection{Minimum Bias}

We examine separately the two contributions to bias. The first contribution,
from effects of finite sampling in the limit $\alpha=1$, is $B_{r,1}=\beta^{T}D_{1}\beta$
following Eq. (\ref{eq:n6}). The weights must satisfy constraints
$u^{T}\beta=1$ and $\beta\geq0$, where $u=\left\{ \begin{array}{ccc}
1 & \ldots & 1\end{array}\right\} ^{T}$, reflecting that weights $\beta_{i}$ sum to 1 and are non-negative.
For minimizing bias, we introduce functional $g_{b,1}\left(\lambda,\beta\right)=\beta^{T}D_{1}\beta+2\lambda\left(1-u^{T}\beta\right)+2\rho^{T}\beta$
that must be stationary at the minimum. The solution must also meet
complementary slackness condition $\rho_{i}\beta_{i}=0$, to account
for inequality constraint $\beta_{i}\geq0$ (\citet{Boyd2004}), and
differentiating $g_{b,1}\left(\lambda,\beta\right)$ yields 
\begin{equation}
D_{1}\beta=\lambda u-\rho\label{eq:n24}
\end{equation}

The next section solves these equations using quadratic programming. 

Here we describe an important factor influencing which observations
receive higher weight. Consider initial guess $x_{0}=\frac{1}{n}u$
for weight-vector $\beta$, corresponding to arithmetic averaging
of the observations. Writing  bias as quadratic function $B_{r,1}\left(x\right)=x^{T}D_{1}x$,
its gradient is $\mathbf{grad}B_{r,1}\left(x\right)=2D_{1}x$. If
one were to revise the vector to $x_{1}=e_{i}$ , the $i$ \textsuperscript{th}
coordinate vector with, for e.g., $e_{1}=\left\{ \begin{array}{ccc}
1 & \ldots & 0\end{array}\right\} ^{T}$, the change is $\triangle x=e_{i}-\frac{1}{n}u$ and the derivative
of $B_{r,1}\left(x\right)$ in the direction of the change is $B_{r,1}'\left(x_{0};\triangle x\right)=\left\{ \mathbf{grad}B_{r,1}\left(x\right)\right\} ^{T}\triangle x/\left\Vert \triangle x\right\Vert $.
Substituting for the gradient, the directional derivative becomes
$B_{r,1}'\left(x_{0};\triangle x\right)=\frac{2}{\left\Vert \triangle x\right\Vert }\left(\frac{1}{n}u^{T}D_{1}e_{i}-\frac{1}{n^{2}}u^{T}D_{1}u\right)$.
This is negative if the $i$ \textsuperscript{th} column average
is smaller than the average across all elements of the matrix $D_{1}$.\footnote{The operation $u^{T}D_{1}u$ calculates the sum of all elements of
$D_{1}$ whereas $u^{T}D_{1}e_{i}$ calculates the sum of elements
of its $i$ \textsuperscript{th} column. } Measurement errors have zero mean so that, for a fixed time $t$:
$\mathbf{E}r_{i}\left(t\right)=\upsilon_{i}\left(t\right)$, and elements
of $D_{1}$ are $D_{1,ij}=\frac{1}{N}\sum_{t}\left(\upsilon_{i}\left(t\right)-\upsilon\left(t\right)\right)\left(\upsilon_{j}\left(t\right)-\upsilon\left(t\right)\right)$.
Hence bias minimization involves generally higher weight to locations
where the expectation is closer to the true spatial average. However,
for observations where this holds the second directional derivative
is generally positive and there is a limit to how far in the direction
$\triangle x/\left\Vert \triangle x\right\Vert $ one can go and still
obtain decreasing $B_{r,1}$. \footnote{With Hessian of $B_{r,1}\left(x\right)$ equal to $2D_{1}$, the second
directional derivative along $\triangle x/\left\Vert \triangle x\right\Vert $
is $B_{r,1}''\left(x_{0};\triangle x\right)=\frac{2}{\left\Vert \triangle x\right\Vert ^{2}}\triangle x^{T}D_{1}\triangle x$,
and substituting yields $B_{r,1}''\left(x_{0};\triangle x\right)=\frac{2}{\left\Vert \triangle x\right\Vert ^{2}}\left\{ \left(e_{i}^{T}D_{1}e_{i}-\frac{1}{n^{2}}u^{T}D_{1}u\right)-\left\Vert \triangle x\right\Vert B_{r,1}'\left(x_{0};\triangle x\right)\right\} .$It
turns out that $\frac{1}{n^{2}}u^{T}D_{1}u\ll e_{i}^{T}D_{1}e_{i}$,
and since $B_{r,1}'\left(x_{0};\triangle x\right)<0$ we have generally
$B_{r,1}''\left(x_{0};\triangle x\right)>0$ so that the directional
derivative is decreasing in magnitude.}

Turning to the second contribution to bias due to missing observations,
we consider the limit where the first contribution is small so that
$\mu_{\upsilon}$ in Eq. (\ref{eq:p22}) approximates $\mathbf{E}\upsilon$.
Then this contribution becomes $B_{r,2}=\left(\frac{1-\alpha}{\alpha}\right)^{2}\mathbf{E}\left\{ \mathbf{E}\upsilon\sum_{i}\beta_{i}^{2}-\sum_{i}\beta_{i}^{2}\mathbf{E}\upsilon_{i}\right\} ^{2}$.
Defining diagonal $n\times n$ matrix $D_{2}\left(t\right)=\left[D_{2,ij}\left(t\right)\right]$
with $D_{2,ii}\left(t\right)=\mathbf{E}\left(\upsilon_{i}-\upsilon\right)$
this contribution to the bias becomes $\left(\frac{1-\alpha}{\alpha}\right)^{2}\left\{ \sum_{i}\beta_{i}^{2}\left(\mathbf{E}\upsilon-\mathbf{E}\upsilon_{i}\right)\right\} ^{2}$,
which is written as $\left(\frac{1-\alpha}{\alpha}\right)^{2}\beta^{T}D_{2}\beta\beta^{T}D_{2}\beta$
and minimizing this requires minimum 
\begin{equation}
\beta^{T}D_{2}\beta\beta^{T}D_{2}\beta\label{eq:n25}
\end{equation}
Introducing functional 
\begin{equation}
g_{b,2}\left(\lambda,\beta\right)=\beta^{T}D_{2}\beta\beta^{T}D_{2}\beta+4\lambda\left(1-u^{T}\beta\right)\label{eq:n26}
\end{equation}
this must be stationary at the optimal solution.\footnote{Although one must also introduce a term $\rho^{T}\beta$ in the functional
along with complementary slackness condition $\rho_{i}\beta_{i}=0$
(\citet{Boyd2004}), to account for inequality constraint $\beta_{i}\geq0$,
we avoid this because, as seen here, the explicit solution to the
equality-constrained problem also meets inequality constraint on weights
$\beta_{i}$. } Differentiating yields cubic polynomials in the $\beta_{i}$s governed
by
\begin{equation}
D_{2}\beta\beta^{T}D_{2}\beta=\lambda u\label{eq:n27}
\end{equation}
using the symmetry of $D_{2}$. If $n=3$ 
\begin{equation}
\left\{ \begin{array}{c}
D_{2,11}^{2}\beta_{1}^{3}+D_{2,11}D_{2,22}\beta_{1}\beta_{2}^{2}+D_{2,11}D_{2,33}\beta_{1}\beta_{3}^{2}\\
D_{2,11}D_{2,22}\beta_{1}^{2}\beta_{2}+D_{2,22}^{2}\beta_{2}^{3}+D_{2,22}D_{2,33}\beta_{2}\beta_{3}^{2}\\
D_{2,11}D_{2,33}\beta_{1}^{2}\beta_{3}+D_{2,22}D_{2,33}\beta_{2}^{2}\beta_{3}+D_{2,33}^{2}\beta_{3}^{3}
\end{array}\right\} =\left\{ \begin{array}{c}
\lambda\\
\lambda\\
\lambda
\end{array}\right\} \label{eq:n28}
\end{equation}
and subtracting each row from the previous one 
\begin{equation}
\left(D_{2,11}\beta_{1}^{2}+D_{2,22}\beta_{2}^{2}+D_{2,33}\beta_{3}^{2}\right)\left(D_{2,11}\beta_{1}-D_{2,22}\beta_{2}\right)=0\label{eq:n29}
\end{equation}
\begin{equation}
\left(D_{2,11}\beta_{1}^{2}+D_{2,22}\beta_{2}^{2}+D_{2,33}\beta_{3}^{2}\right)\left(D_{2,22}\beta_{2}-D_{2,33}\beta_{3}\right)=0\label{eq:n30}
\end{equation}
Consider the case $D_{2,11}<0$ and $D_{2,22},D_{2,33}>0$, so that
the first location has expectation larger than the long-term spatial
average, whereas the others have smaller expected values. Then $D_{2,22}\beta_{2}-D_{2,33}\beta_{3}=0$
or 
\begin{equation}
\frac{\beta_{2}}{\beta_{3}}=\frac{\mathbf{\mathbf{E}\upsilon_{3}-E}\upsilon}{\mathbf{\mathbf{E}\upsilon_{2}-E}\upsilon}\label{eq:n31}
\end{equation}
Higher weight is given to locations with expectation closer to the
long-term spatial average, although the precise relationship is different
from minimizing $B_{r,1}$. Additionally $D_{2,11}\beta_{1}^{2}+D_{2,22}\beta_{2}^{2}+D_{2,33}\beta_{3}^{2}=0$
or 
\begin{equation}
\frac{\beta_{1}}{\beta_{2}}=\left\{ -\frac{D_{2,22}}{D_{2,11}}\left(1+\frac{D_{2,22}}{D_{2,33}}\right)\right\} ^{1/2}\label{eq:n32}
\end{equation}
In case $D_{2,11}=-2$, $D_{2,22}=1$ and $D_{2,33}=1$, then $\beta_{1}=\beta_{2}=\beta_{3}=\frac{1}{3}$.
If $D_{2,11}=-3$, $D_{2,22}=1$ and $D_{2,33}=3$, then $\beta_{1}=\frac{1}{3}$,
$\beta_{2}=\frac{1}{2}$ and $\beta_{3}=\frac{1}{6}$. Generally the
minimizing of $B_{r,2}$ requires higher weights for locations where
the value of the field is expected to be closer to the long-term spatial
average. In the extreme case with one observation having $D_{2,ii}=0$
we obtain simply $\beta_{i}=1$. If some location has the same expectation
as the spatial average then only it needs to be sampled in order to
minimize the bias due to missing observations.

For general $n$ 
\begin{equation}
\left\{ \begin{array}{c}
D_{2,11}\beta_{1}\sum_{k=1}^{n}D_{2,kk}\beta_{k}^{2}\\
\ldots\\
\ldots\\
D_{2,nn}\beta_{n}\sum_{k=1}^{n}D_{2,kk}\beta_{k}^{2}
\end{array}\right\} =\left\{ \begin{array}{c}
\lambda\\
\ldots\\
\ldots\\
\lambda
\end{array}\right\} \label{eq:p33}
\end{equation}
and subtracting each row from the previous one 
\begin{equation}
\left\{ \begin{array}{c}
\left(D_{2,11}\beta_{1}-D_{2,22}\beta_{2}\right)\sum_{k=1}^{n}D_{2,kk}\beta_{k}^{2}\\
\ldots\\
\left(D_{2,n-1,n-1}\beta_{n-1}-D_{2,nn}\beta_{n}\right)\sum_{k=1}^{n}D_{2,kk}\beta_{k}^{2}
\end{array}\right\} =\left\{ \begin{array}{c}
0\\
\ldots\\
0
\end{array}\right\} \label{eq:p34}
\end{equation}
which is solved by first considering the independent relations among
the two groups of locations having $D_{2,ii}>0$ and $D_{2,ii}<0$,
for a total of $n-2$ equations, and then solving the remaining two
equations $\sum_{k=1}^{n}D_{2,kk}\beta_{k}^{2}=0$ and $\sum_{k=1}^{n}\beta_{k}=1$.
The  bias increases with $\left(\frac{1-\alpha}{\alpha}\right)^{2}$. 

In summary, minimizing bias due to finite sampling as well as from
missing observations involves larger weights to locations where the
field lies closer to the true spatial average, although the precise
models are different. This is hardly surprising, because bias from
finite sampling depends on a non-diagonal matrix with elements $\mathbf{E}\left(\upsilon_{i}\left(t\right)-\upsilon\left(t\right)\right)\left(\upsilon_{j}\left(t\right)-\upsilon\left(t\right)\right)$,
whereas that from missing observations depends on a diagonal matrix
having elements $\mathbf{E}\left(\upsilon_{i}-\upsilon\right)$. 

\subsection{Minimum Variance}

As discussed in the previous section, in case $\alpha<1$ the variance
of the spatial average is generally not quadratic in the weights.
However, even in this case we find it instructive to imagine the limiting
case of small $B_{r,1}$ so that $\sum_{i}\beta_{i}\mathbf{E}\upsilon_{i}\cong\mathbf{E}\upsilon$,
which is a constant, so that the variance of the spatial average simplifies
to quadratic form 
\begin{equation}
\sigma_{r}^{2}=\beta^{T}C\beta\label{eq:p35}
\end{equation}
with
\begin{equation}
C=C_{1}+C_{2}+C_{3}+C_{4}+C_{5}\label{eq:p36}
\end{equation}
being the sum of $n\times n$ symmetric matrices describing respective
terms in Eq.  (\ref{eq:p23}). The optimal weights minimize functional
\begin{equation}
g_{\upsilon}\left(\lambda,\beta\right)=\beta^{T}C\beta+2\lambda\left(1-u^{T}\beta\right)+2\rho^{T}\beta\label{eq:p37}
\end{equation}
along with complementary slackness condition $\rho_{i}\beta_{i}=0$,
to account for inequality constraint $\beta_{i}\geq0$ and for stationarity
\begin{equation}
C\beta=\lambda u-\rho\label{eq:p38}
\end{equation}
Let us consider some special cases to develop intuition.

\subsubsection{Observations are always available}

In case $\alpha=1$ then $C_{4}=C_{5}=0$ and $\varsigma_{\upsilon_{i}}^{2}=\sigma_{\upsilon_{i}}^{2}$
so that $C_{1}+C_{2}=S_{\upsilon}$, the covariance matrix of field
$\upsilon$, and $C_{3}=\sigma_{\varepsilon}^{2}I$, where $I$ is
the identity matrix. Then $C=S_{\upsilon}+\sigma_{\varepsilon}^{2}I$.
This corresponds to the formula found by previous authors who assumed
that observations can be counted on being available (\citet{Gandin1993,Vinnikov1999}). 

In the limit $\sigma_{\varepsilon}^{2}\rightarrow0$ if observations
are precise, $C=S_{\upsilon}$ and $\beta=P_{\upsilon}\left(\lambda u-\rho\right)$,
where $P_{\upsilon}=S_{\upsilon}^{-1}$ is the inverse covariance,
or precision, matrix of the field. In a field with zero spatial correlation
so that $S_{\upsilon}$ is a diagonal matrix containing terms $\sigma_{\upsilon_{i}}^{2}$
then the precision matrix is also diagonal and weight $\beta_{i}$
is proportional to $1/\sigma_{\upsilon_{i}}^{2}$, being higher for
locations where the variance is smaller. 

The opposite extreme where measurement uncertainty is so large that
$C\cong\sigma_{\varepsilon}^{2}I$ leads to uniform weights $\beta_{i}=1/n$. 

Generally the weights must take into account both the precision matrix
of the field and measurement variance. Writing $Q_{\upsilon}=\frac{1}{\sigma_{\varepsilon}^{2}}S_{\upsilon}$,
$C^{-1}=\frac{1}{\sigma_{\varepsilon}^{2}}\left(I+Q_{v}\right)^{-1}$,
which is equal to $\frac{1}{\sigma_{\varepsilon}^{2}}\left\{ I-\left(I+Q_{\upsilon}^{-1}\right)^{-1}\right\} $.\footnote{We have used identity involving matrices $U,$ $V$, $W$, $Z$ $\left(U+WVZ\right)^{-1}=U^{-1}-U^{-1}W\left(V^{-1}+ZU^{-1}W\right)^{-1}ZU^{-1}$,
and set $W=Z=I$ (\citet{Zhang1999}).} In case diagonal elements of $Q_{\upsilon}^{-1}$ are much larger
than one, corresponding to diagonal elements of the precision matrix
being much larger than $1/\sigma_{\varepsilon}^{2}$, one can approximate
$C^{-1}\cong\frac{1}{\sigma_{\varepsilon}^{2}}\left(I-Q_{\upsilon}\right)$.
This is a small perturbation to the case of large measurement uncertainty,
and the weights are proportional to $\frac{1}{\sigma_{\varepsilon}^{2}}\left(I-Q_{\upsilon}\right)u$. 

The general case needs to be considered numerically, but its interpretation
is quite simple. Notice that $S_{\upsilon}+\sigma_{\varepsilon}^{2}I$
is $S_{r}$, the covariance matrix of observations $r_{i}$, because
the observation error is assumed to be independent of true value $\upsilon_{i}$
and errors are independent of each other. 

\subsubsection{Field is spatially uncorrelated}

Let us reconsider the case where the field is spatially uncorrelated,
but where any measurement goes unrecorded with probability $1-\alpha$.
Then $C$ is a diagonal matrix with 
\begin{equation}
c_{ii}=\frac{1}{\alpha}\varsigma_{\upsilon_{i}}^{2}+\frac{1}{\alpha}\sigma_{\varepsilon}^{2}-2\frac{1-\alpha}{\alpha}\upsilon\mathbf{E}\upsilon_{i}+\frac{1-\alpha}{\alpha}\upsilon^{2}\label{eq:p39}
\end{equation}
and, using the expression for $\varsigma_{\upsilon_{i}}$ in Eq.  (\ref{eq:n20})
\begin{equation}
c_{ii}=\frac{1}{\alpha}\left(\sigma_{\upsilon_{i}}^{2}+\sigma_{\varepsilon}^{2}\right)+\frac{1-\alpha}{\alpha}\left(\upsilon-\mathbf{E}\upsilon_{i}\right)^{2}\label{eq:p40}
\end{equation}
and $\beta_{i}\propto1/c_{ii}$. Higher weights are given to locations
with lower variance and those with expectation closer to the spatial
mean. The second factor becomes more important if the probability
of missing observations is higher. In the general case this is modified
to account for effects of spatial covariance through the precision
matrix. 

\subsubsection{General case}

From the previous development the general case can be denoted as minimizing
$\sigma_{r}^{2}=\beta^{T}C\beta$ where 
\begin{equation}
C=\frac{1}{\alpha}S_{r}+\frac{1-\alpha}{\alpha}F_{\upsilon}\label{eq:p41}
\end{equation}
where, as noted earlier, $S_{r}$ is the covariance matrix of observations
and $F_{\upsilon}$ is a diagonal matrix with $i$\textsuperscript{th}
diagonal entry $\left(\mathbf{E}\upsilon_{i}-\mathbf{E}\upsilon\right)^{2}$.
The second term is similar to the contribution of missing data to
bias, except for the form of dependence on $\mathbf{E}\upsilon_{i}-\mathbf{E}\upsilon$.
Minimizing contributions to bias and variance from missing observations
both require emphasizing in some manner observations with expected
value near the spatial average. The variance is inversely proportional
to probability $\alpha$ of reporting individual observations, and
in the limit $\alpha\rightarrow0$ the variance becomes infinite. 

\subsection{Minimum Mean squared error (MSE) through Quadratic Programming}

If $\alpha<1$ neither bias nor variance is quadratic in weights $\beta_{i}$.
In case the first contribution to bias from finite sampling is much
larger, as it can be expected to be if $\alpha$ is closer to 1, we
may approximate bias as a quadratic form in $\beta$. Second, if the
contribution to variance from missing data is small compared to intrinsic
variability, then variance too can be approximated as a quadratic
form by assuming $\sum_{i}\beta_{i}\mathbf{E}\upsilon_{i}\cong\mathbf{E}\upsilon$.
As a result the MSE becomes quadratic in the weights: $\beta^{T}\left(C+D_{1}\right)\beta$,
and the following section chooses optimal weights using quadratic
programming to minimize this quantity. Of course, owing to the simplifications
made, this is only an approximate minimum. However, once minimizing
solutions are found, bias and variance can be computed more accurately
for the corresponding averaging scheme, from Eqs. (\ref{eq:p22})
and (\ref{eq:p23}). 

\section{Computational results}

We apply these developments to gridded rain-gauge data products covering
the Indian mainland, released by India Meteorological Department (IMD)
(\citet{Rajeevan2006,Pai2014}) at scales of $1^{\circ}\times1^{\circ}$
(\citet{Rajeevan2005,Rajeevan2006}) and subsequently at higher resolution
of $0.25^{\circ}\times0.25^{\circ}$ using a much larger network of
rain-gauge stations (\citet{Pai2014}). These datasets were prepared
by interpolating data from individual rain gauges onto a regular grid,
with weights inversely depending on squared distance to grid\textquoteright s
midpoints (\citet{Rajeevan2005}), following the scheme of \citet{Shepard1968}.
The higher resolution dataset is found comparable to the previous
gridded rainfall datasets in many aspects but furthermore elicits
more accurately the rainfall amounts in regions exhibiting larger
spatial gradients (\citet{Pai2014}). Therefore in the present paper
we use the $0.25^{\circ}\times0.25^{\circ}$ dataset for the months
of April-November to estimate area-averaged rainfall, treating it
as true rainfall, and analyze time-series from the $1^{\circ}\times1^{\circ}$
dataset as if they represented a sparser sample of $357$ distinct
daily observations, from which the spatial average is to be estimated. 

Bias and variance are modeled in Section 2 by heavily truncating a
Taylor series. The number of terms in the Taylor series of $f\left(R,S\right)$
describing a ratio is infinite, and the \char`\"{}delta-method\char`\"{}
used here would converge only if all the moments of $R$ and $S$
defined in Eq. (\ref{eq:n1}), appearing as they do in progressively
higher terms in the series, were finite. Figure 1 considers the quality
of the resulting approximation of squared bias and variance by comparing
with simulations. Graphs marked \char`\"{}simulations\char`\"{} have
been computed as follows: for each period we simulate each of the
$357$ values of $s_{i}$, i.e. availability of individual observations,
as independent Bernoulli random variables with $s_{i}=1$ occurring
with probability $\alpha$. Once the $s_{i}$'s are known for a given
year, Indian Summer Monsoon Rainfall (ISMR) is estimated from Eq.
(\ref{eq:n1}) for that year, where $r_{i}$ corresponds to average
rainfall from June through September.\footnote{We happen to choose weights $\beta_{i}$ to minimize MSE, but an alternative
choice could well have been made for this figure, given our goal here
of validating models of bias and variance.} Repeating this process, for each of the years 1901-2011, yields a
single time-series for ISMR. We simulate $5000$ such realizations
of the time-series of ISMR, reflecting uncertainty in which observations
are reported, and compute the ensemble mean ISMR as the mean across
realizations. The simulated bias is computed from the ensemble averaged
time-series of ISMR, whereas the simulated variance is the average
of the temporal variance of each realization.

We compare with the squared bias and variance estimated from models
in Eqs. (\ref{eq:p22}) and (\ref{eq:p23}). Figure 1 shows that these
models, based on truncating the respective series, perform rather
well, diverging from simulations only for very small $\alpha$. Even
for an extreme case of $90$ \% probability of individual point observations
not being reported, the models reach within $10$ \% of simulated
bias and variance for these datasets. In realistic applications, we
can expect much smaller probabilities of missing data, with $\alpha$
being close to $1$, so that the models of Eqs. (\ref{eq:p22})-(\ref{eq:p23})
should perform adequately in case of variables for which the series
converges similarly to that of rainfall. 

Another important illustration from these plots is an important effect
that missing data has in terms of increasing the bias and variance
in the spatial average. The effects are substantial if availability
of individual observations is small. However, in the more relevant
case where $\alpha$ is closer to $1$, the bias and variance are
dominated by intrinsic features. In particular, bias in estimates
of the spatial average is dominated by effects of finite sampling
of a continuous field and variance is mainly from inherent variability
in the process and, to a much lesser degree, effects of measurement
noise. Even so, the effects of missing data can generally not be neglected. 

Figure 2 and 3 plot optimal weights for minimizing squared bias, variance,
and MSE. A basic difficulty in analyzing bias and MSE for continuous
fields is that the true value averaged over time and space is generally
unknown. Rain gauges provide continuous measurements at what are approximately
points and satellites, while providing a larger field of view, offer
only brief snapshots in time (\citet{Bell2003}). For want of a better
alternative, we treat results from IMD's $0.25^{\circ}\times0.25^{\circ}$
gridded rainfall dataset (\citet{Pai2014}) as approximating the true
ISMR upon being area-averaged. We then consider values from the $1^{\circ}\times1^{\circ}$
dataset (\citet{Rajeevan2006}) as yielding individual point-observations
that must be weighted. As for availability of observations, we only
consider cases nearer to the more realistic regime, in which $\alpha$
is close to $1$, and this furthermore permits us to use quadratic
programming to approximate the optimal weights for minimizing variance
and MSE (Section 3.3). This is possible because, with $\alpha$ closer
to $1$, the contribution of missing observations to bias and variance
is relatively small, as Figure 1 shows, and optimal weights can be
approximated by neglecting terms higher than quadratic in the $\beta_{i}$s.
However, in case of bias, the contribution of missing data cannot
be simplified through a quadratic function in the weights, and we
limit analysis to $\alpha=1$. 

A striking feature of OA schemes (Figures 2-3) is that only a small
fraction of potential observations is needed for estimating the spatial
average. As described in Section 3, minimizing bias generally requires
giving higher weights to locations where rainfall has expectation
closer to the spatial average (Figure 2b), whereas reducing variance
requires higher weights to locations with small variance and covariance.
If $\alpha$ is smaller, more locations need to be included in the
OA scheme. Bias and variance both contribute significantly to MSE,
and hence its minimization includes features of reducing both bias
and variance. 

Figure 4a shows the time-series of ISMR obtained by area-averaging
the $0.25^{\circ}\times0.25^{\circ}$ dataset, which we treat as the
true values. Also shown are ensemble averages for the MSE-minimizing
scheme in cases of $\alpha=1.0$ and $\alpha=0.8$. Ensembles in this
case consider different realizations of $s_{i}$ for each location
and year and, of course, if $\alpha=1.0$ then all realizations are
the same. Lower availability $\alpha$ leads to higher temporal variance,
and the graph illustrates that this is manifested through overestimating,
compared to $\alpha=1$, in years of above-average ISMR and underestimating
in other years. 

Figure 4b-d show cumulative frequency distributions of the weights
for the three different OA schemes. The first bin, involving smallest
$\beta_{i}$'s, has been omitted because it comprises mainly zero
weights describing locations not appearing in the OA scheme. Therefore
the lowest ordinates in the curves of Figure 4b-d indicate approximately
what fraction of the overall domain is not involved in the corresponding
OA scheme, this generally being quite large. With smaller availability
of individual observations, more of the domain participates in the
OA scheme, with weights becoming slightly more evenly distributed. 

We compute standard error (SE) in the spatial average. The SE measures
uncertainty in reports of the all-India average, and is computed as
the square root of the contribution to MSE from measurement uncertainty
and squared bias as $SE=\left\{ \beta^{T}\left(D_{1}+\sigma_{\varepsilon}^{2}I\right)\beta\right\} ^{1/2}$
. Contribution of measurement uncertainty is $\sigma_{\varepsilon}^{2}\beta^{T}\beta$,
and we assume all observations being reported, so $\alpha=1$, in
which case bias reduces to $\beta^{T}D_{1}\beta$. Weights $\beta$
are chosen so as to minimize MSE, and are shown in Figure 3b.\footnote{For uniform weights $\beta=\frac{1}{n}u$ the SE would $\frac{1}{n}\left\{ u^{T}\left(D_{1}+\sigma_{\varepsilon}^{2}I\right)u\right\} ^{1/2}$}
Figure 5 graphs the SE associated with estimates of all-India rainfall
for individual months between April and November. We consider two
different cases of measurement uncertainty depicted by $\sigma_{e}$.
The small difference between the two cases shows that propagation
of this uncertainty is limited by the spatial averaging process even
when only a small fraction of the domain is involved in the OA scheme,
so that the standard error has contribution mainly from bias. The
present analysis assumes $\alpha=1$, so bias is the result of using
point measurements to estimate a spatial average. 

The OA scheme used to estimate the standard error minimizes MSE in
the spatial average. Instead, we might have chosen to minimize bias
directly, but that would yield a time-series with much larger variance
and thereby higher MSE. MSE-minimizing schemes must generally be chosen
over bias-minimizing schemes if we seek to compare differences between
the estimate and the true value, because bias involves only the expected
value of the estimate. Such schemes that minimize the MSE lead to
irreducible standard error if there is intrinsic variability in the
field. Figure 5 also plots standard deviation (Stdev) of the optimal
average as well as its long-term mean. The SE, while being smaller
than standard deviation, is substantial. 

For ISMR, averaged from June-September, after accounting for reductions
in $\sigma_{e}$ from monthly to the 4-monthly time-scale, the SE
is approximately $0.29$ mm/day, mean is $7$ mm/day and standard
deviation is $0.39$ mm/day. 

To examine robustness of these results to assumptions about true ISMR,
we repeat the analysis with true spatial average $\upsilon$ being
estimated by area-weighting the $1^{\circ}\times1^{\circ}$ (\citet{Rajeevan2005,Rajeevan2006})
dataset, while continuing to use the same dataset in the OA scheme
(Figure 6). Bias minimization recovers weights generally increasing
with the cosine of latitude, while variance minimization yields identical
results as before (Figure 3a), since variance does not depend on the
true spatial average. The main result is that even MSE minimization
yields a similar averaging scheme (compare Figures 3b and 6c). In
the presence of variance, the OA scheme that minimizes MSE does not
appear sensitive to the choice of dataset that represents true values
of ISMR. Figure 6d shows that the standard error exhibits similar
variation across months as in Figure 5, and numerical values are comparable
to the previous analysis. The ratio of SE / mean ISMR for the two
analyses is $4.1$ \% and $4.3$ \% respectively. 

\begin{figure}
\includegraphics[scale=0.9]{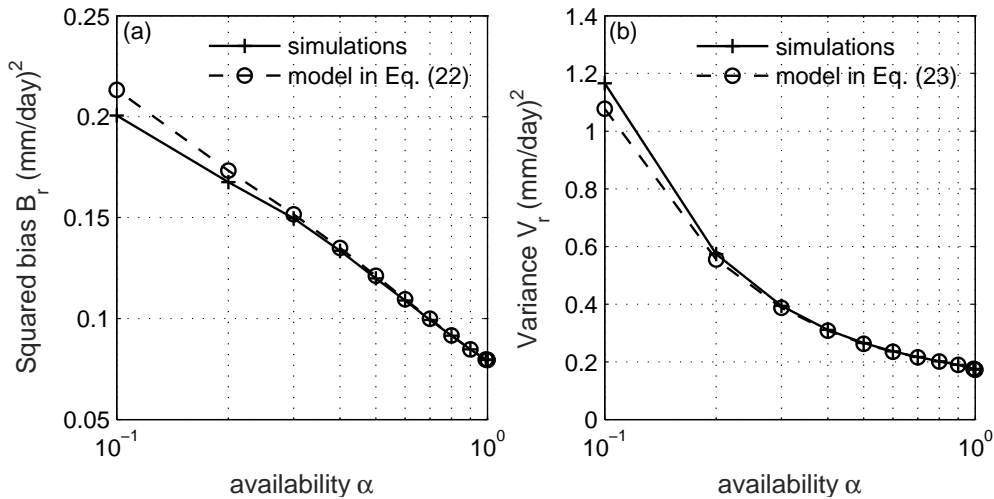}

\caption{Verification of models of bias $B_{r}$ and variance $V_{r}$ of a
spatial average (Indian Summer Monsoon Rainfall, ISMR) in Eqs. (\ref{eq:p22})-(\ref{eq:p23})
for different values of availability $\alpha$, by comparison with
simulations. Generally, except for small $\alpha$, corresponding
to high probability of missing data, these models based on truncation
of Taylor series provide good approximations. }
\end{figure}

\begin{figure}
\includegraphics[scale=0.8]{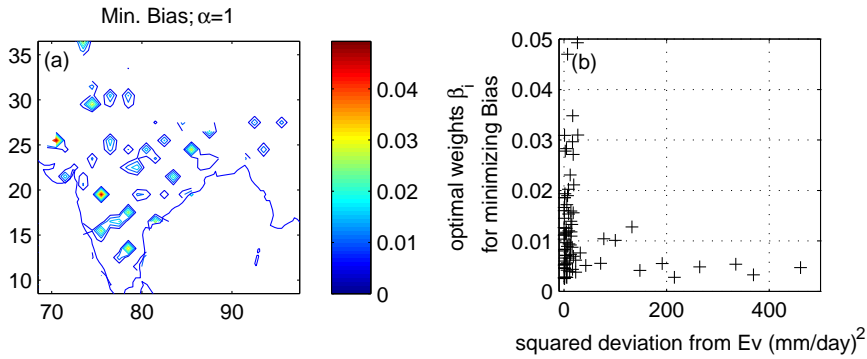}

\caption{Optimal weights for minimizing squared bias in estimate of ISMR and
graphed as a function of $\left(\mathbf{E}\upsilon-\mathbf{E}\upsilon_{i}\right)^{2}$.
Large weights occur only at locations where the deviation from the
spatial average is small. }
\end{figure}

\begin{figure}
\includegraphics{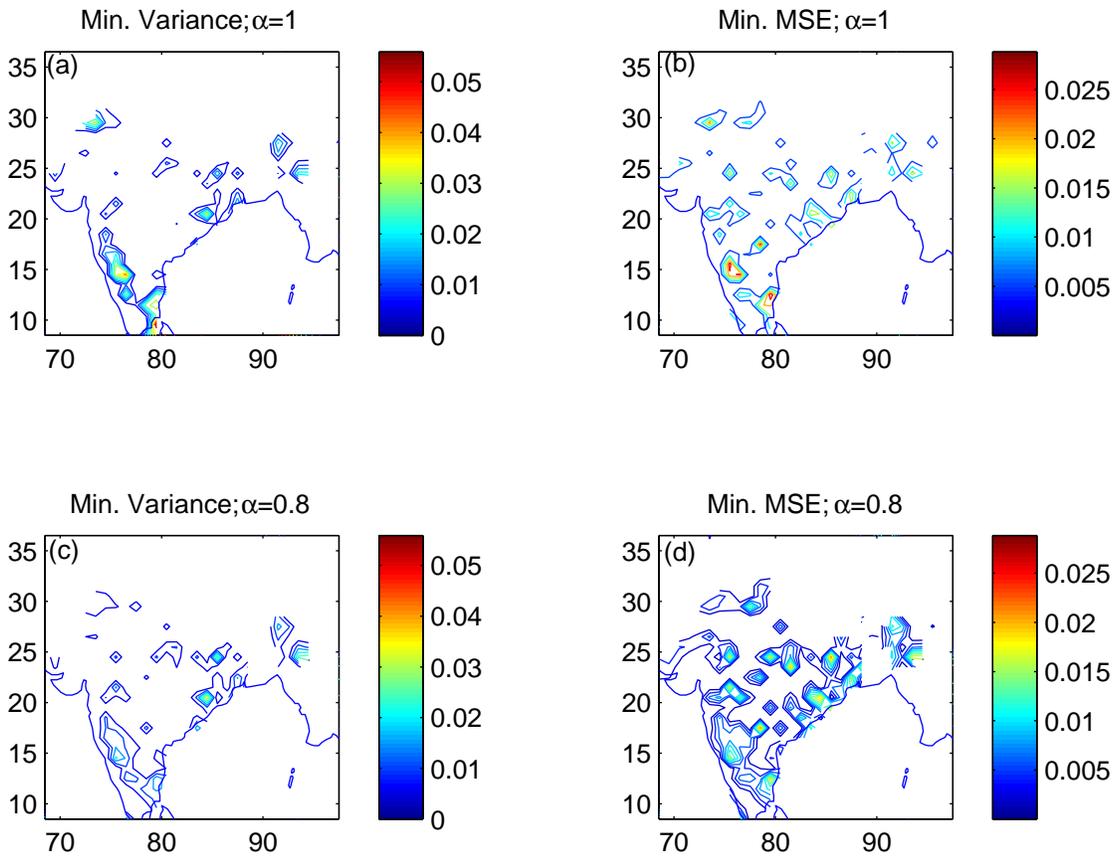}

\caption{Optimal weights for minimizing variance and mean squared error of
ISMR, in case $\alpha=1.0$ (upper panels) and $\alpha=0.8$ (lower
panels).}
\end{figure}

\begin{figure}
\includegraphics[scale=0.8]{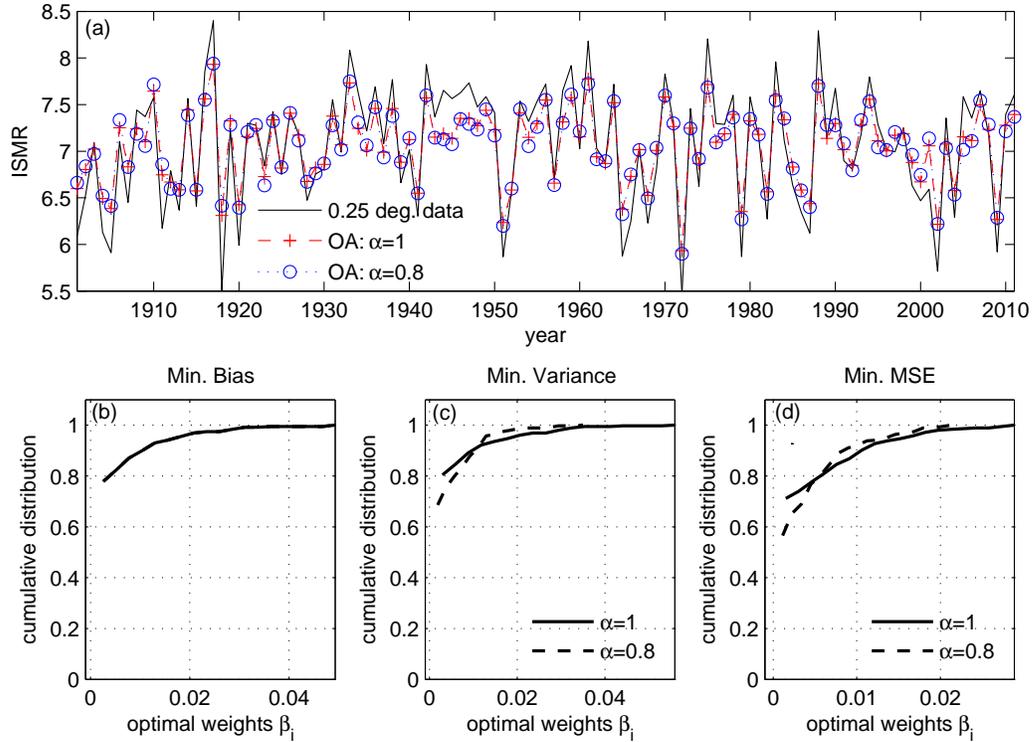}

\caption{(a) Time-series of ISMR obtained by area-averaging IMD's gridded rainfall
dataset at $0.25^{\circ}\times0.25^{\circ}$ resolution, and results
of optimal averaging (OA) (for minimum-MSE) the dataset at $1^{\circ}\times1{}^{\circ}$
resolution in case $\alpha=1.0$ and $\alpha=0.8$ ; (b)-(d) Cumulative
distribution of optimal weights $\beta_{i}$ for minimizing bias,
variance and MSE respectively. The first bin (i.e. smallest $\beta_{i}s$)
is not plotted, because it mostly comprises weights that are zero.
Therefore the lowest ordinate of the curves indicates the fraction
of area that need not be included in OA. For e.g. bias minimization
requires sampling about $20$\% of total area. Minimizing variance
or MSE requires more of the area to be sampled if $\alpha$ is smaller
and weights are then distributed more evenly across locations.   }

\end{figure}
\begin{figure}
\includegraphics[scale=0.8]{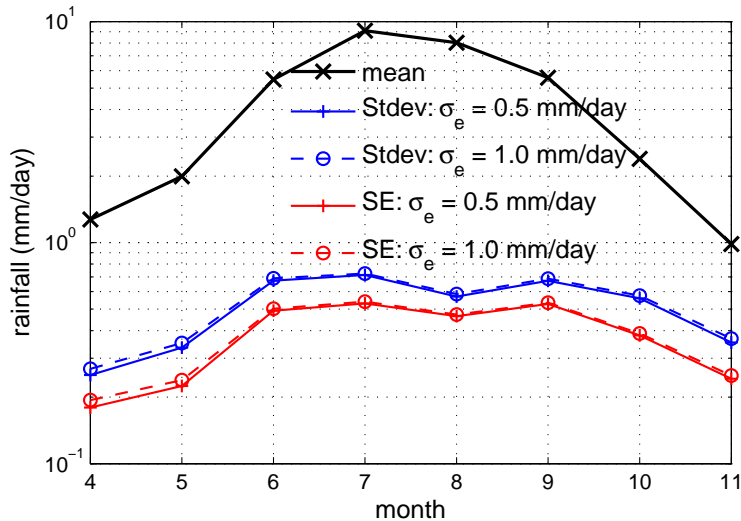}

\caption{Estimated standard error (SE) of the OA scheme that minimizes MSE,
for individual months ranging from April through November. The formula
used for calculating standard error is $SE=\left\{ \beta^{T}\left(D_{1}+\sigma_{\varepsilon}^{2}I\right)\beta\right\} ^{1/2}$,
where weights $\beta$ are chosen to minimize MSE. Also shown are
standard deviation and mean rainfall. Measurement error has little
effect and the standard error comes mainly from bias. For ISMR (June
- September) the standard error is approximately $0.29$ mm/day, mean
is $7$ mm/day and standard deviation $0.39$ mm/day. }
\end{figure}

\begin{figure}
\includegraphics[scale=0.8]{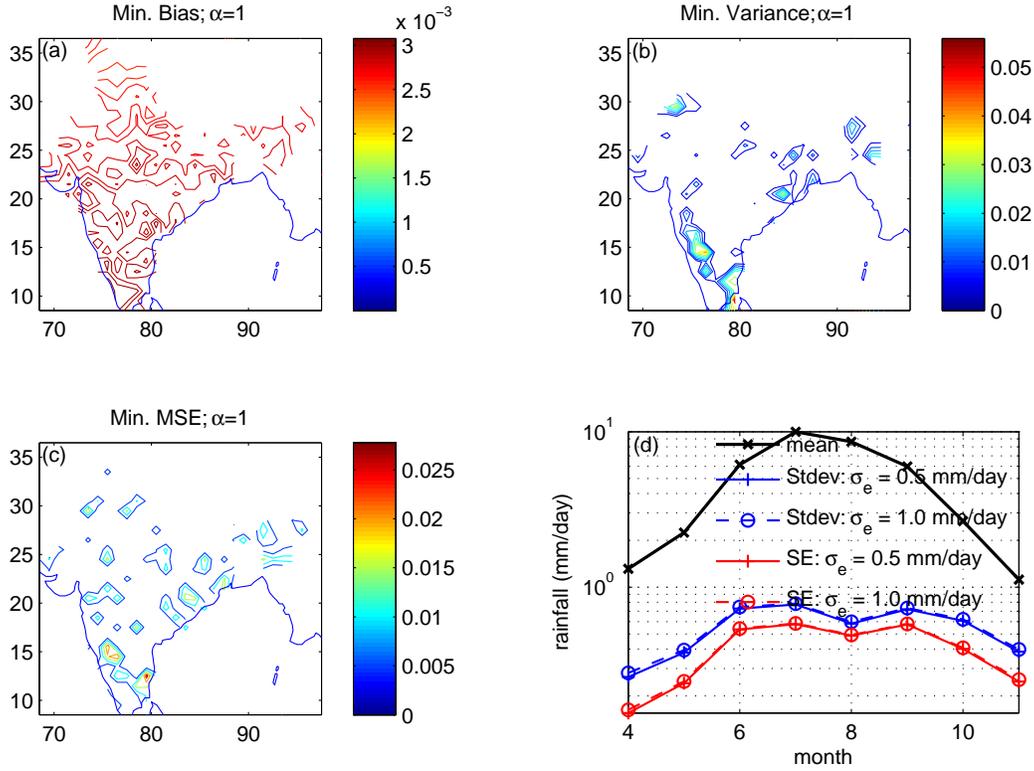}

\caption{Analysis repeated after treating area-averaged rainfall from $1^{\circ}\times1^{\circ}$
rain gauge data as true values. Shown are: (a)-(c) optimal weights
for minimizing bias, variance, and MSE in estimates of ISMR, for $\alpha=1$,
and (d) estimated SE of the OA scheme that minimizes MSE, for individual
months ranging from April through November. Bias minimization recovers
weights that are approximately increasing in cosine of latitude. Optimal
weights for minimizing variance are identical to results in Figure
3a (where $0.25^{\circ}\times0.25^{\circ}$ dataset was treated as
true values). Optimal weights for minimizing MSE are similar to the
results in Figure 3b. Properties of standard error in Figure 5 are
similar to the present analysis in Figure 6d. Here, for ISMR the SE
is approximately $0.33$ mm/day, mean is $7.7$ mm/day and standard
deviation $0.43$ mm/day. }
\end{figure}

\pagebreak{}

\section{Discussion}

Spatial averages appear in derived climate variables such as global
mean surface temperature and regional rainfall (\citet{Mooley1984,Vinnikov1990}).
We estimate bias, variance, and mean squared error of a ratio describing
a spatial average, in the presence of missing observations. The numerator
is linearly weighted point observations over some spatial domain and
the denominator represents the fraction of unity represented by available
observations. The \char`\"{}delta-method\char`\"{} (\citet{Oehlert1992,Vaart1998})
derives estimators by taking expectations of truncated Taylor series
of the ratio. The resulting estimators are non-parametric, with no
assumptions being made about the distribution of the underlying variable.
However, imagine a sequence of approximations to the bias and variance
of a ratio, involving successive terms in the respective Taylor series.
These successive terms involve progressively higher moments of the
numerator and denominator, so convergence requires the moments to
be finite (\citet{Oehlert1992}). 

The estimators for bias and variance of a spatial average were tested
on gridded rain-gauge data over India (\citet{Rajeevan2006,Pai2014}).
The models, based on truncating the respective series to $1$\textsuperscript{st}
order for variance and $2$\textsuperscript{nd} order for bias, fare
very well, diverging from simulations only for very small $\alpha$,
the probability of individual observations being reported. Whether
or not such simplified approximations would converge for a different
variable depends on behavior of the successive moments for the temporal
and spatial scales of averaging. 

Missing observations increase bias and variance, with effects increasing
with the probability that individual observations are missing. Expressions
developed here describe prior estimates of bias and variance of the
sample spatial average obtained from identical measuring devices.
Such prior estimates are useful when it is not known which of the
potential point observations would actually be available in any particular
instance. Of course, the incorporation of knowledge of which observations
are actually available in any given case would affect particular estimates
of bias and variance. 

We also examine optimal weights that minimize bias, variance, or MSE.
Previous authors (\citet{Vinnikov1990,Gandin1993,Shen1998,Vinnikov1999})
have considered this problem and the present work is an extension
to include effects of missing observations. In the present analysis
both the numerator and denominator of the ratio describing a spatial
average are random variables. The present estimators reduce to the
previous results (\citet{Gandin1993,Zhang1999}) for a special case:
describing that either the probability of missing data is zero or
all observations are found to have been recorded, both of which are
statistically indistinguishable. 

Analysis of optimal sampling subsumes the problem of choosing how
to locate measurements in order to minimize bias, variance, or a combination
of the two as well as the problem of how to use existing observations.
If a system of measurements is in place, the choice of weights can
help reveal how to use potential observations in a weighting scheme
whose goal is to minimize either the bias, variance or MSE. Such has
been the motivation of prior discussions of optimal weighting for
climate data (\citet{Vinnikov1990,Gandin1993,Shen1998,Vinnikov1999}).
In addition, optimal weights can also reveal how to situate measuring
devices. 

Generally, optimal weighting procedures require to sample only a small
fraction of the total area, but this fraction increases in the presence
of missing observations. Minimizing variance of the spatial average
requires giving higher weights to locations having smaller variance
and covariance. The possibility of missing records introduces an additional
factor, arising from the squared difference between the expected value
at the location and the spatial mean, whose importance increases in
proportion to the probability of missing data. Therefore the possibility
of missing data generally affects how weights should be chosen in
order to minimize variance. Bias, both due to finite sampling and
that due to missing data, can be minimized by emphasizing locations
where the expectation is closer the the spatial average. 

We computed standard error (SE) in estimates of all-India rainfall
for individual months between April and November, and for monsoon
rainfall between June - September (ISMR). For this analysis, we used
a $0.25^{\circ}\times0.25^{\circ}$ gridded rain-gauge product\citet{Pai2014})
to estimate area-averaged rainfall, treating it as true rainfall,
and treated time-series from a $1^{\circ}\times1^{\circ}$ gridded
rain-gauge product (\citet{Rajeevan2005,Rajeevan2006}) as observations
in the OA procedure. The SE characterizes uncertainty in reports of
the all-India average, and includes contributions to the MSE from
measurement uncertainty as well as the squared bias. Because individual
measurements are modeled as having noise with zero mean, the effect
of measurement uncertainty at the all-India scale is reduced, and
the SE mainly comes from bias due to finite sampling of a continuous
field. Our analysis assumed that area-averages from gridded rain-gauge
data at $0.25^{\circ}\times0.25^{\circ}$ resolution come closest
to the true values, which is unknown in practice. 

Repeating the analysis with true spatial average estimated by area-weighting
the $1^{\circ}\times1^{\circ}$ (\citet{Rajeevan2005,Rajeevan2006})
dataset yields similar results for minimum-MSE weights and the standard
error, providing a measure of confidence in this result. Due to spatiotemporal
variability, the optimal averaging scheme that minimizes MSE is not
sensitive to assumptions about the true value of ISMR, and hence it
appears that the SE can be estimated. Its value is about 4 \% of mean
ISMR, and this uncertainty in estimation should be considered in statistical
inference problems involving all-India rainfall. 

\section*{Acknowledgments}

This work has been supported by Divecha Centre for Climate Change,
Indian Institute of Science. Thanks to colleagues for helpful suggestions.
Code and data used for generating the figures is in Supplementary
Information.

\section*{Appendix 1: Sufficient conditions for validity of variance estimator}

Our approximation of the variance of a ratio in Eq. (\ref{eq:n7})
assumed $\boldsymbol{\mathbf{E}}f\left(R,S\right)\cong\frac{\boldsymbol{\mathbf{E}}R}{\boldsymbol{\mathbf{E}}S}$,
which from Eq. (\ref{eq:p3}) requires
\begin{equation}
\frac{\boldsymbol{\mathbf{E}}\left(S-\boldsymbol{\mathbf{E}}S\right)^{2}\frac{\boldsymbol{\mathbf{E}}R}{\left(\boldsymbol{\mathbf{E}}S\right)^{3}}-\boldsymbol{\mathbf{E}}\left(R-\boldsymbol{\mathbf{E}}R\right)\left(S-\boldsymbol{\mathbf{E}}S\right)\frac{1}{\left(\boldsymbol{\mathbf{E}}S\right)^{2}}}{\frac{\boldsymbol{\mathbf{E}}R}{\boldsymbol{\mathbf{E}}S}}\label{eq:p43}
\end{equation}
to be negligible compared to unity. Equivalently $\frac{\sigma_{S}^{2}}{\mu_{S}^{2}}-\frac{\sigma_{RS}^{2}}{\mu_{R}\mu_{S}}=\left(\frac{1-\alpha}{\alpha}\right)\left\{ \sum_{i}\beta_{i}^{2}-\frac{\sum_{i}\beta_{i}^{2}\mathbf{E}\upsilon_{i}}{\sum_{i}\beta_{i}\mathbf{E}\upsilon_{i}}\right\} $
must be small, for which it is sufficient that $1-\alpha\ll1$ . However
this is not necessary, since $\sum_{i}\beta_{i}=1$, and therefore
generally $\sum_{i}\beta_{i}^{2}\ll1$. 

\section*{Appendix 2: Variance and covariance of $R$ and $S$}

The variance of $R$ is
\begin{equation}
\sigma_{R}^{2}\equiv\boldsymbol{\mathbf{E}}\left(R-\mathbf{E}R\right)^{2}=\mathbf{E}R^{2}-\left(\mathbf{E}R\right)^{2}\label{eq:p44}
\end{equation}
with first term expanding to
\begin{equation}
\mathbf{E}R^{2}=\mathbf{E}\sum_{i}r_{i}^{2}a_{i}^{2}s_{i}^{2}+2\mathbf{E}\sum_{i<j}r_{i}r_{j}a_{i}a_{j}s_{i}s_{j}\label{eq:p45}
\end{equation}
and, using linearity of $\mathbf{E}$, independence between observed
value $r_{i}$ and availability $s_{i}$, and independence between
$s_{i}$ and $s_{j}$ for $i\neq j$ 
\begin{equation}
\mathbf{E}R^{2}=\alpha\sum_{i}a_{i}^{2}\mathbf{E}r_{i}^{2}+2\alpha^{2}\sum_{i<j}a_{i}a_{j}\mathbf{E}r_{i}r_{j}\label{eq:p46}
\end{equation}
using $\mathbf{E}s_{i}^{2}=\alpha$ and, if $i\neq j$, $\mathbf{E}s_{i}s_{j}=\mathbf{E}s_{i}\mathbf{E}s_{j}=\alpha^{2}$.
The second term in Eq.  (\ref{eq:p44}), using linearity of $\mathbf{E}$,
expands to
\begin{equation}
\left(\mathbf{E}R\right)^{2}=\sum_{i}\left(\mathbf{E}r_{i}a_{i}s_{i}\right)^{2}+2\sum_{i<j}\mathbf{E}r_{i}a_{i}s_{i}\mathbf{E}r_{j}a_{j}s_{j}\label{eq:p47}
\end{equation}
and, using independence between $r_{i}$ and $s_{i}$
\begin{equation}
\left(\mathbf{E}R\right)^{2}=\alpha^{2}\sum_{i}a_{i}^{2}\left(\mathbf{E}r_{i}\right)^{2}+2\alpha^{2}\sum_{i<j}a_{i}a_{j}\mathbf{E}r_{i}\mathbf{E}r_{j}\label{eq:p48}
\end{equation}
using $\mathbf{E}s_{i}=\alpha$ . Hence
\begin{equation}
\sigma_{R}^{2}=\alpha\sum_{i}a_{i}^{2}\left\{ \mathbf{E}r_{i}^{2}-\alpha\left(\mathbf{E}r_{i}\right)^{2}\right\} +2\alpha^{2}\sum_{i<j}a_{i}a_{j}\mathbf{Cov}\left(r_{i},r_{j}\right)\label{eq:p49}
\end{equation}
where $\mathbf{Cov}\left(r_{i},r_{j}\right)=\mathbf{E}r_{i}r_{j}-\mathbf{E}r_{i}\mathbf{E}r_{j}$. 

The variance of $S$ is 
\begin{equation}
\sigma_{S}^{2}\equiv\boldsymbol{\mathbf{E}}\left(S-\mathbf{E}S\right)^{2}=\mathbf{E}S^{2}-\left(\mathbf{E}S\right)^{2}\label{eq:p50}
\end{equation}
whose first term simplifies to 
\begin{equation}
\mathbf{E}S^{2}=\alpha\sum_{i}a_{i}^{2}+2\alpha^{2}\sum_{i<j}a_{i}a_{j}\label{eq:p51}
\end{equation}
using linearity of $\mathbf{E}$, independence of $s_{i}$ and $s_{j}$,
and $\mathbf{E}s_{i}=\alpha$. Similarly the second term in Eq.  (\ref{eq:p50})
simplifies to
\begin{equation}
\left(\mathbf{E}S\right)^{2}=\alpha^{2}\sum_{i}a_{i}^{2}+2\alpha^{2}\sum_{i<j}a_{i}a_{j}\label{eq:p52}
\end{equation}

Hence the variance of $S$ is
\begin{equation}
\sigma_{S}^{2}=\alpha\left(1-\alpha\right)\sum_{i}a_{i}^{2}\label{eq:p53}
\end{equation}
As for covariance between $R$ and $S$
\begin{equation}
\sigma_{RS}^{2}\equiv\mathbf{Cov}\left(R,S\right)=\mathbf{E}RS-\mathbf{E}R\mathbf{E}S\label{eq:p54}
\end{equation}
whose first term, using linearity of $\mathbf{E}$, becomes 
\begin{equation}
\mathbf{E}RS=\sum_{i}\mathbf{E}r_{i}a_{i}^{2}s_{i}^{2}+2\sum_{i<j}\mathbf{E}r_{i}a_{i}a_{j}s_{i}s_{j}\label{eq:p55}
\end{equation}
simplifying to
\begin{equation}
\alpha\sum_{i}a_{i}^{2}\mathbf{E}r_{i}+2\alpha^{2}\sum_{i<j}a_{i}a_{j}\mathbf{E}r_{i}\label{eq:p56}
\end{equation}
using independence of $r_{i}$ and $s_{i}$ and of $s_{i}$ and $s_{j}$.
The second term in Eq.  (\ref{eq:p54}) becomes 
\begin{equation}
\mathbf{E}R\mathbf{E}S=\sum_{i}\mathbf{E}r_{i}a_{i}s_{i}\mathbf{E}a_{i}s_{i}+2\sum_{i<j}\mathbf{E}r_{i}a_{i}s_{i}\mathbf{E}a_{j}s_{j}\label{eq:p57}
\end{equation}
simplifying to
\begin{equation}
\alpha^{2}\sum_{i}a_{i}^{2}\mathbf{E}r_{i}+2\alpha^{2}\sum_{i<j}a_{i}a_{j}\mathbf{E}r_{i}\label{eq:p58}
\end{equation}
using independence of $r_{i}$ and $s_{i}$. 

Therefore covariance between $R$ and $S$ is 
\begin{equation}
\sigma_{RS}^{2}=\alpha\left(1-\alpha\right)\sum_{i}a_{i}^{2}\mathbf{E}r_{i}\label{eq:p59}
\end{equation}
\pagebreak{}

\bibliographystyle{plainnat}
\bibliography{Spatial_average_refs}

\end{document}